# Temperature-dependent Raman scattering of $DyScO_3$ and $GdScO_3$ single crystals

O Chaix-Pluchery[1], D Sauer and J Kreisel

*Laboratoire des Matériaux et du Génie Physique, CNRS UMR 5628, Grenoble Institute of Technology, Minatec, 3, parvis Louis Néel, 38016 Grenoble, France*

**Abstract**

We report a temperature-dependent Raman scattering investigation of $DyScO_3$ and $GdScO_3$ single crystals from room temperature up to 1200 °C. With increasing temperature, all modes decrease monotonously in wavenumber without anomaly, which attests the absence of a structural phase transition. The high temperature spectral signature and extrapolation of band positions to higher temperatures suggest a decreasing orthorhombic distortion towards the ideal cubic structure. Our study indicates that this orthorhombic-to-cubic phase transition is close to or higher than the melting point of both rare-earth scandates ($\cong$ 2100 °C), which might exclude the possibility of the experimental observation of such a phase transition before melting. The temperature-dependent shift of Raman phonons is also discussed in the context of thermal expansion.

[1] Author to whom any correspondence should be addressed (*E-mail: Odette.Chaix@grenoble-inp.fr*)

## 1. Introduction

The understanding of $ABO_3$ perovskite-type oxides is a very active research area with great relevance to both fundamental- and application-related issues, particularly related with dielectric and ferroelectric properties [1, 2]. The ideal cubic structure of $ABO_3$ perovskites is rather simple, with corner-linked anion octahedra $BO_6$, the $B$ cations at the centre of the octahedra and the $A$ cations in the space between the octahedra. With respect to this ideal cubic perovskite, structural distortions can be described by separating two main features [3-5]: (*i*) a rotation (tilt) of essentially rigid $BO_6$ octahedra and/or (*ii*) polar cation displacements. Past investigations of these instabilities have been a rich source for the understanding of structural properties not only in perovskites but also in oxides in general.

In this paper we will focus on perovskites which present no cation displacement but only the anti-ferrodistortive (AFD) tilt distortion. It is well known [6] that this tilt distortion (the tilt angle) can be driven by external parameters such as temperature or pressure as exemplified in the model materials $SrTiO_3$ (STO) [7, 8] or $LaAlO_3$ (LAO) [9, 10]. While it has been shown that *pressure* can either increase [11] or reduce [10, 12, 13] AFD-instabilities at the zone-boundary, it is generally accepted that *temperature* reduces tilt instabilities, i.e. the tilt angle away from the cubic structure decreases with increasing temperature. However, the critical temperature $T_c$ at which the structure becomes cubic can greatly vary (e.g. $T_{c,STO}$ = -163°C and $T_{c,LAO}$ = 527°C) [7, 9] and some tilted perovskites decrease in tilt angle but do not become cubic before their melting.

Our present work focuses on two rare-earth scandates (*RE*-scandates), $DyScO_3$ and $GdScO_3$. They adopt an orthorhombic structure (*Pnma* space group), which derives from the ideal perovskite structure by an anti-phase tilting of the adjacent $ScO_6$ octahedra ($a^-b^+a^-$ in Glazer's notation [3, 4]). *RE*-scandates receive currently an active research interest as a potential high-k dielectric material [14-19] and as a substrate material for the epitaxial growth of strained high-quality perovskite-type thin films [20-25]. Beyond the well adapted unit cell parameters for growth of several perovskite oxides, the crystalline perfection and homogeneity of the $DyScO_3$ and $GdScO_3$ single crystals and the rather low dislocation density make scandate single crystals some of the best available substrates for the epitaxial growth of perovskite thin films [20]. Given the fact that thin film growth takes places at high temperature, information on the temperature dependent behaviour such as phase transitions or thermal expansion is of importance. A previous X-ray diffraction (XRD) study indicates no phase transition in the 25 to 1000 °C range, though the orthorhombicity decreases with increasing temperature [26]. It is however well known that subtle phase transitions in perovskites are often below XRD resolution or very difficult to be detected [5, 27]. The thermal expansion coefficients of $DyScO_3$ and $GdScO_3$ were determined in this temperature range [26] but the phonon thermal behaviour, and thus the Grüneisen parameters, remain unknown.

In the following we will present an investigation of $DyScO_3$ and $GdScO_3$ single crystals by temperature-dependent Raman scattering. Raman spectroscopy, which probes zone-centre phonons, is known to be a versatile technique for the investigation of oxide materials in particular for the detection of small structural distortions in perovskites [28-34], even when they are too subtle to be detected by diffraction techniques [30, 35]. Building up on our previous room temperature Raman study of DyScO$_3$ and GdScO$_3$ [36], the aim of our study will be the investigation of the phonon spectrum up to 1200 °C, namely in view of a possible phase transition. To the best of our knowledge, this is the first temperature-dependent phonon investigations of any *RE*-scandate.

## 2. Experimental

DyScO$_3$ and GdScO$_3$ single crystal platelets supplied by *CrysTec* [37] have been investigated by Raman spectroscopy up to 1200 °C, using a commercial Linkam heating stage (TS1500) placed under the Raman microscope. Raman spectra were collected using a Jobin Yvon/Horiba LabRam spectrometer equipped with a liquid nitrogen cooled charge coupled device detector. Experiments were conducted in the micro-Raman mode in a backscattering geometry; the instrumental resolution was 2.8 ± 0.2 cm$^{-1}$. The 514.5 nm line of an Ar$^+$ ion laser was used as excitation line to record DyScO$_3$ spectra. For the investigation of GdScO$_3$ we used the 632.8 nm line of an He-Ne laser in order to limit the recently reported fluorescence, which almost entirely masks the phonon signature [36]. Experiments were carried out with a laser power below 5 mW on the sample. The light was focused to a 1 μm$^2$ spot using a 50lf objective. Spectra were recorded every 50 °C in the range from room temperature up to 1200°C for DyScO$_3$ and 1050°C for GdScO$_3$ with an acquisition time varying between 50 and 100 sec. Spectra were calibrated using the Si spectrum at room temperature. The Raman spectra before and after heating are identical, attesting the reversibility of temperature-induced changes. The fitting procedure of the Raman lines was performed using the Labspec software of Horiba/Jobin Yvon by using Lorentzian profiles after subtraction of the baseline.

## 3. Results and discussion

### 3. 1. Raman spectra

Before presenting the experimental Raman investigation, we remind that the primitive unit cell of the cubic perovskite contains one formula unit *ABX*$_3$ giving rise to 15 vibrational modes at the zone centre, which decompose into the following optical phonons: $\Gamma_{Pm-3m}$ (opt) = 3 F$_{1u}$ + F$_{2u}$, where F$_{1u}$ modes are IR-active and F$_{2u}$ is a silent mode and none is Raman active. On the other hand, the four *RE*ScO$_3$ formula units (thus 20 atoms) per unit cell in the orthorhombic *Pnma* structure, give rise to 60 vibrational modes at the zone centre, as predicted from factor group analysis: $\Gamma_{Pnma}$(opt) = 7 A$_g$ + 5 B$_{1g}$ + 7 B$_{2g}$ + 5 B$_{3g}$ + 8 A$_u$ + 9 B$_{1u}$ + 7 B$_{2u}$ + 9 B$_{3u}$, whereof 24 modes are Raman-active (7 A$_g$ + 5 B$_{1g}$ + 7 B$_{2g}$ + 5 B$_{3g}$), 25 modes are IR-active (9 B$_{1u}$ + 7 B$_{2u}$ + 9 B$_{3u}$) and 8 modes are silent (8 A$_u$). In a previous work we have reported polarized Raman spectra at room temperature, leading to the assignments of 23 modes out of the 24 predicted modes [36]. The experimental band positions for DyScO$_3$ are for most bands in reasonable agreement with recent theoretical *ab initio* predictions of the vibrational spectrum for the same material [18].

The here presented temperature-dependent Raman experiments on DyScO$_3$ and GdScO$_3$ have been carried out in two different configurations (see [36] for notations with respect to crystal orientations): In a first experiment, spectra have been recorded without polarization analysis, this configuration led to spectra which are rather similar to those obtained in the parallel polarization configuration *y(x'x')y* or *y(z'z')y* in the pseudo-cubic reference setting and contain mainly $A_g$ and $B_{2g}$ modes. A second experiment has been carried out in a configuration of parallel (VV) polarizer and analyser [*y(xx)y* or *y(zz)y* in the orthorhombic reference setting, i.e. after 45° rotation of the crystal platelets]. In this parallel configuration only modes of $A_g$ symmetry are expected. Figures 1a and 1b present Raman spectra obtained at room temperature for DyScO$_3$ and GdScO$_3$, respectively, in both configurations of polarisation. Although the VV spectra are very similar to our previously reported data [36], we observe, further to the predicted intense $A_g$ modes, some weak modes of $B_{2g}$ and $B_{1g}$ symmetry also, which is likely due to a slight sample disorientation.

Temperature-dependent Raman spectra of DyScO$_3$ and GdScO$_3$ are presented in figures 2a-2b and in figure 3. It can be seen that the temperature evolution of depolarized phonon spectra is rather similar for both compounds. At high temperature the spectra are first disturbed and then masked by an increasing thermal black body radiation, which has inhibited measurements above 1200 °C. As expected, the Raman spectra of GdScO$_3$ which are recorded using a 633 nm excitation are earlier disturbed (starting at 750°C) than the DyScO$_3$ spectra recorded at 514 nm (starting at 950°C). Further to this, the very high temperature GdScO$_3$ spectra are disturbed by unexplained optical interference fringes below 240 cm$^{-1}$. As expected, all Raman lines shift towards low wavenumbers and broaden with increasing temperature. The fact that the spectral signature evolves continuously without suppression or appearance of new bands suggests that DyScO$_3$ and GdScO$_3$ undergo no phase transition in the investigated temperature range.

Although not all bands can be fitted up to the highest temperature, mainly due to band broadening and in turn band overlap, most bands can be reliably fitted. Figures 4a and 4b present the quantitative temperature evolution of the band positions for DyScO$_3$ and GdScO$_3$, respectively. The position of modes collected in the two different polarization configurations are very close, as illustrated in figure 5 for the three $A_g$ modes at high wavenumber. With increasing temperature, all modes decrease monotonously in wavenumber without any anomaly which again attests the absence of a structural phase transition. As seen in Table 1, the slopes from a linear curve fit for each mode are rather similar in both compounds. The modes can be classified into three types depending upon the slope values. $A_g$ modes observed below 300 cm$^{-1}$ shift slowly (-0.15 to -1.05 cm$^{-1}$/100°C) whereas, above 300 cm$^{-1}$, the shift is much faster (-2.1 to -3.3 cm$^{-1}$/100°C). The $B_{1g}$(2) mode measured up to 900°C for GdScO$_3$ shifts slowly (-0.59 cm$^{-1}$/100°C) similar to the $A_g$ modes below 300 cm$^{-1}$ and to the 176 cm$^{-1}$ mode in DyScO$_3$ (mode not assigned in ref. 36). B$_{2g}$ modes are situated in an intermediate zone (-1.2 to -2.0 cm$^{-1}$/100°C).

It is interesting to note the very high temperature spectra of DyScO$_3$ and GdScO$_3$ can be qualitatively described by three broad features, although the fine structure modes are still needed for the fit up to the highest temperature. The three broad features are approximately centred at 142, 301, 441 cm$^{-1}$ and 145, 288, 439 cm$^{-1}$ for DyScO$_3$ and GdScO$_3$, respectively. Generally speaking, the tendency to reduce the number of bands points to an evolution towards a higher symmetry. The fact that we observe three large phonon features is reminiscent of the three expected infrared-active phonon modes in the ideal cubic *Pm-3m* perovskite structure. While it is common to observe three bands in infrared spectra for a structure that approaches a metrically cubic cell [38], it is more surprising to observe such features in Raman spectroscopy. Nevertheless, we take the signature of the three dominating Raman features as a strong argument for a decreasing orthorhombic distortion with a structural evolution of DyScO$_3$ and GdScO$_3$ towards the cubic *Pm-3m* perovskite structure at high temperature. We then interpret the high temperature Raman signature as being dominated by thermal disorder (breaking the strict Raman selection rule). This tendency towards a cubic structure based on spectroscopic arguments is in good agreement with an earlier reported thermal expansion study by XRD of DyScO$_3$ and GdScO$_3$ [26]. In order to estimate the critical temperature at which an orthorhombic-to-cubic phase transition could take place, we have extrapolated the linear fit of the modes to higher temperatures to identify their point of intersection. Results (not shown here) indicate that three couples of A$_g$-B$_{2g}$ modes intersect at very high temperatures. The lowest intersection temperature is found at approximately 2035°C for B$_{2g}$(5) and A$_g$(7) in DyScO$_3$ and 2320°C for A$_g$(2) and B$_{2g}$(2) in GdScO$_3$. These temperatures are close to or higher than the melting point of both rare-earth scandates ($\cong$ 2100°C [20, 39]), which might exclude the possibility of the experimental observation of such a phase transition before melting.

*3.2 Thermal expansion*

The thermal expansion coefficients of DyScO$_3$ and GdScO$_3$ have been determined earlier from 25 to 1000°C using XRD [26]. The calculated volumetric thermal expansion coefficients of DyScO$_3$ and GdScO$_3$ are $\alpha_{V,DSO}$ = 2.42 x 10$^{-5}$ K$^{-1}$ and $\alpha_{V,GSO}$ = 3.07 x 10$^{-5}$ K$^{-1}$, respectively.

We have seen in the previous section that the frequency of all Raman modes decreases to a good approximation linearly with temperature from room temperature to 1200 °C. Such a linear temperature dependence makes the Raman modes themselves in principle suitable for the determination of the thermal expansion, assuming that changes in the volume *ΔV* and thus in bond distances *Δd*, induce changes in force constants and thus in vibrational frequencies. The Grüneisen model assumes that the Grüneisen constant *γ* correlates the temperature dependence of vibrational frequencies with the unit cell volume. At constant pressure the Grüneisen constant is defined by

$$\gamma = \left.\frac{\partial \ln \nu}{\partial \ln V}\right|_P = -\frac{1}{\alpha_V}\left.\frac{\partial \ln \nu}{\partial T}\right|_P \quad (1)$$

where $\nu$ is the frequency of a vibrational mode, $V$ the unit cell volume and $\alpha_V$ the volumetric thermal expansion given by

$$\alpha_V = \frac{1}{V}\left.\frac{\partial V}{\partial T}\right|_p \quad (2)$$

As a consequence, when the temperature dependence of vibrational modes is known, the Grüneisen parameter can be determined from the known thermal expansion (or *vice-versa*). The so-calculated Grüneisen parameters are given in Table 1 under the assumption that the same volumetric thermal expansion $\alpha_V$ can be used for every mode. As expected the Grüneisen parameters vary significantly from one mode to the other.

It is interesting to remind that the $\alpha_V$-values in ref. [26] are based on the approximation [40] that $\alpha_V$ of a tilt-distorted $ABO_3$ perovskite can be empirically expressed as a decoupled combination of changes of the $B$-O bond length (thermal expansion $\alpha_{V,oct}$ of regular octahedra) and the tilting angle of $BO_6$ octahedral framework (thermal expansion $\alpha_\phi$) according to $\alpha_V = \alpha_{V,oct} + \alpha_\phi$. The thermal expansion is thus calculated by considering only the $BO_6$ octahedra while the A-O bond is taken into account indirectly via $\alpha_\phi$. We parenthesize that such an approach is similar to the so-called polyhedral approach, based on comparative crystal chemistry and developed by Hazen and Finger which has been shown to be useful in predicting elastic properties of a large number of solids [41, 42].

Based on the principle of such a polyhedra-type approach, it has been proposed that elastic properties such as thermal expansion or compressibility of perovskites can be estimated through the analysis of particular Raman modes, even without explicit knowledge of the Grüneisen parameters [43-45]. In the following we will test this model by using our temperature-dependent Raman spectra of $DyScO_3$ and $GdScO_3$. The method described in detail in ref. [43] is based on the following assumptions. (*i*) Changes in the lattice parameter $a$ induce changes in force constants $k$ and consequently in vibrational frequencies (Grüneisen model) and (*ii*) polyhedron stretching modes involve mainly one type of force constants and the thermal expansion (or compressibility) of the polyhedron is determined by these microscopic force constants. Based on these assumptions, the approximation of a pseudo-cubic structure and that the force constants obey an empiric law for ionic crystals as described in ref. [46], Loridant et al. [43] have proposed that the thermal expansion of a polyhedron can be described by the following empirical equation:

$$\alpha_V = \frac{1}{V}\left.\frac{\partial V}{\partial T}\right|_p \cong -\frac{6}{13}\frac{1}{\nu}\frac{\Delta \nu}{\Delta T} \quad (3)$$

For testing this surprisingly simple empirical law we need to identify vibrational modes which are representative for the thermal expansion. The authors of [43] have suggested to identify modes being representative of both $BO_6$ and $AO_{12}$ polyhedra, which are the constituent polyhedra of perovskites; it is however not straightforward in which way (parallel circuit or serial circuit) the polyhedral elastic values should be summed up [43, 45, 47]. Following the discussion above and the approximations in [26, 40], we will consider only one mode which is characteristic of the $ScO_6$ octahedra vibrations: the high wavenumber $A_{1g}(7)$ in-phase stretching mode of $ScO_6$ octahedra (assignment from [48]). As a stretching mode, this mode involves directly Sc-O force constants and its $\Delta\nu/\Delta T$ slope is characteristic of the response of the elastic system to the external parameter temperature. Furthermore, it has been shown [49] that the position in wavenumber of this mode is for orthorhombic *Pnma* perovskites directly related to the octahedra tilt angle according to the empirical relationship 23.5 cm$^{-1}$/deg. For the 509 cm$^{-1}$ mode of $DyScO_3$ this relationship gives an octahedra tilt angle of 22° in good agreement with the value of 21° determined from XRD [50].

Applying equation (3) to the $A_{1g}(7)$ mode leads to $\alpha_{V,DSO} = 2.99 \times 10^{-5}$ K$^{-1}$ and $\alpha_{V, GSO} = 2.85 \times 10^{-5}$ K$^{-1}$ which have to be compared to the XRD determined values [26] $\alpha_{V,DSO} = 2.42 \times 10^{-5}$ K$^{-1}$ and $\alpha_{V, GSO} = 3.07 \times 10^{-5}$ K$^{-1}$. The good agreement suggests that the model and approximations of ref. [43], but also the chosen mode, provide a valuable approach for a first rough estimation of the thermal expansion via Raman modes, namely for materials for which experimental XRD thermal expansion data is not reported. However, more work is needed for further testing (and eventually improving) this simple Raman-mode-based model.

## 4. Conclusion

We have presented a temperature-dependent Raman scattering investigation of $DyScO_3$ and $GdScO_3$ single crystals from room temperature up to 1200 °C. Within this temperature range the crystals undergo no structural phase transition as attested by the fact that all modes decrease monotonously in wavenumber without anomaly. The spectral evolution suggests a decreasing orthorhombic distortion towards the ideal cubic structure. However, a potential orthorhombic-to-cubic phase transition is close to, or higher than, the melting point of both rare-earth scandates ($\cong$ 2100 °C), which might exclude the possibility of the experimental observation of such a phase transition before melting. It is shown that the temperature-dependent shift of a specific Raman phonons can be used for a rough estimation of the thermal expansion, thus validating the model and approximations of ref. [43].

*Acknowledgements*
This work was supported by the European Strep MaCoMuFi and the French National Founding agency (ANR) within the "Proper" project.

**Figure captions**

**Figure 1:** Raman spectra of (a) DyScO$_3$ and (b) GdScO$_3$ in both the unpolarized configuration and the parallel (VV) polarization configuration [orthorhombic reference setting (y(xx)y or y(zz)y]. The A$_g$ mode positions are given in bold.

**Figure 2**: Temperature-dependent depolarized Raman spectra of (a) DyScO$_3$ and (b) GdScO$_3$.

**Figure 3**: Temperature-dependent VV-polarized Raman spectra of DyScO$_3$. The intensity of the spectrum at 1050°C is multiplied by a factor of 5.

**Figure 4**: Temperature dependence of the positions of (a) DyScO$_3$ and (b) GdScO$_3$ for the well-defined Raman modes. The lines are linear curve fits in the investigated temperature range.

**Figure 5**: Temperature dependence of the positions of the three A$_g$ modes at high wavenumber of DyScO$_3$ and GdScO$_3$ collected in the two different polarization configurations (full symbols for polarized spectra, open symbols for depolarized spectra).

**Table 1:** Positions in wavenumber, slope values of linear curve fits in figure 4 and Grüneisen parameter γ of each well-defined Raman mode in $DyScO_3$ and $GdScO_3$. The Grüneisen constants were derived from equation (1) in the text, by using the volumetric thermal expansion values $\alpha_V$ given in ref. [26] ($\alpha_V = 2.42 \times 10^{-5}$ $K^{-1}$ for $DyScO_3$ and $\alpha_V = 3.07 \times 10^{-5}$ $K^{-1}$ for $GdScO_3$).

| Mode assignment | $DyScO_3$ | | | $GdScO_3$ | | |
|---|---|---|---|---|---|---|
| | ω (cm⁻¹) at 20°C | δω/δT (cm⁻¹/100°C) | γ | ω (cm⁻¹) at 20°C | δω/δT (cm⁻¹/100°C) | γ |
| $A_g(1)$ | 111 | -0.87 | 3.24 | 113 | -1.05 | 3.03 |
| $A_g(2)$ | 131 | -0.19 | 0.60 | 131 | -0.15 | 0.37 |
| $B_{2g}(2)$ | 156 | -1.3 | 3.44 | 155 | -1.2 | 2.52 |
| ? | 176 | -0.89 | 2.09 | - | - | - |
| $B_{1g}(2)$ | - | - | - | 223 | -0.59 | 0.86 |
| $A_g(3)$ | 254 | -0.91 | 1.48 | 248 | -0.63 | 0.83 |
| $B_{2g}(3)$ | 309 | -1.7 | 2.27 | 298 | -1.7 | 1.86 |
| $A_g(4)$ | 327 | -2.2 | 2.78 | 321 | -2.3 | 2.33 |
| $B_{2g}(4)$ | 355 | -2.0 | 2.33 | 351 | -1.9 | 1.76 |
| $A_g(5)$ | 434 | -2.9 | 2.76 | 418 | -2.8 | 2.18 |
| $A_g(6)$ | 459 | -2.1 | 1.89 | 452 | -2.1 | 1.51 |
| $B_{2g}(5)$ | 475 | -1.6 | 1.39 | 463 | -1.5 | 1.06 |
| $A_g(7)$ | 509 | -3.3 | 2.68 | 501 | -3.1 | 2.02 |
| $B_{2g}(6)$ | 542 | -2.0 | 1.53 | 532 | -1.7 | 1.04 |

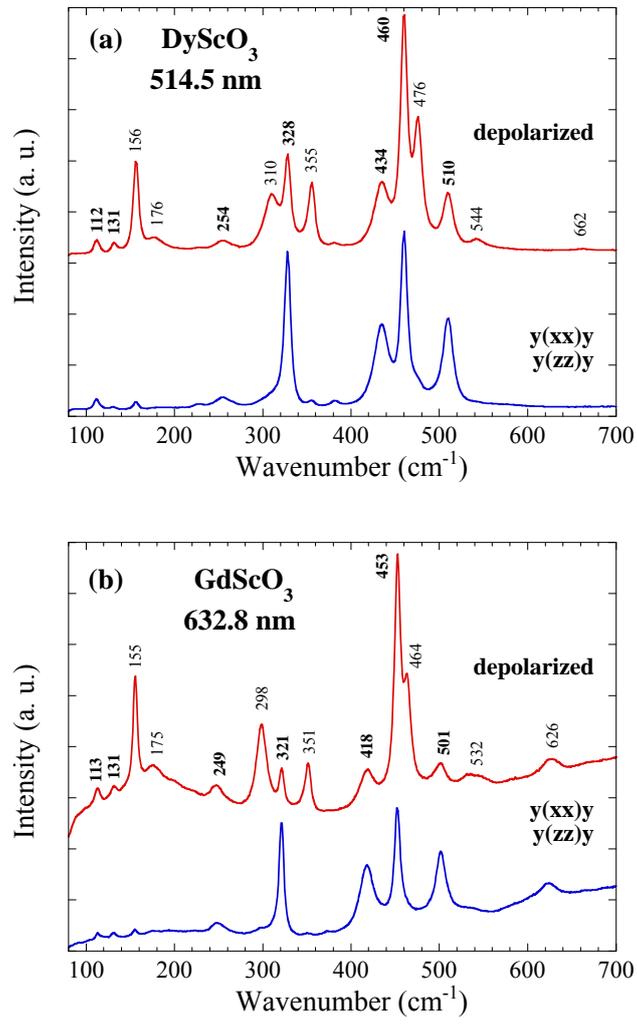

**Figure 1**

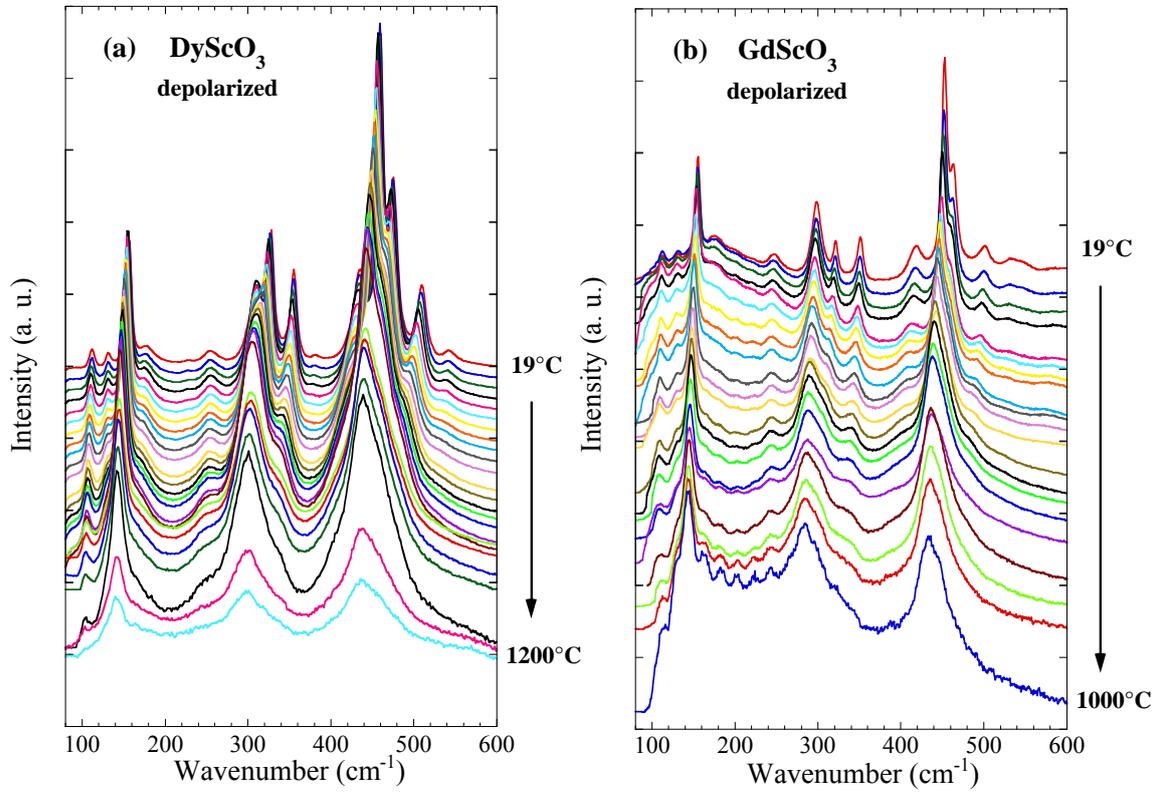

**Figure 2**

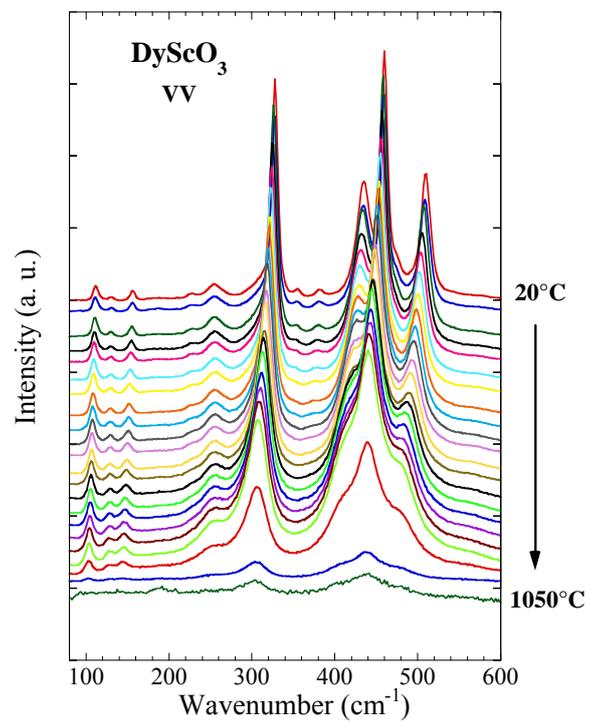

**Figure 3**

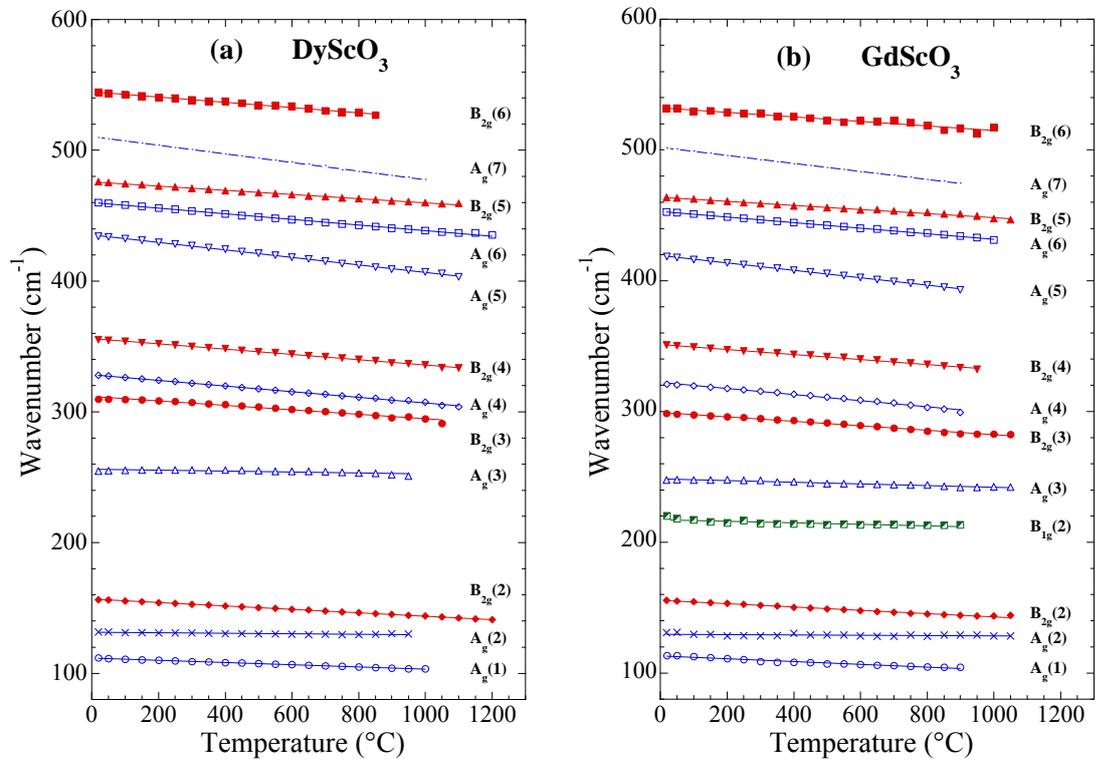

**Figure 4**

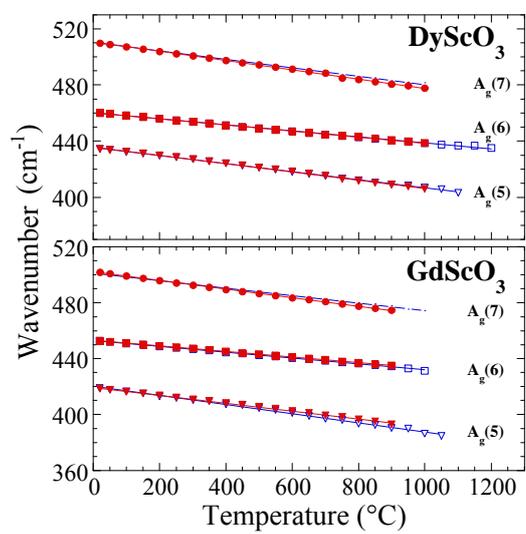

**Figure 5**